\documentclass[11pt,twoside]{article}

\usepackage{asp2006}
\usepackage{epsf}
\usepackage{graphicx}
\usepackage{lscape}

\begin{document}
\title{Bars in Starbursts and AGNs -- A Quantitative Reexamination}   
\author{Lei Hao,$^{1}$ Shardha Jogee,$^{1}$ Fabio D. Barazza,$^{2}$ Irina Marinova,$^{1}$ and Juntai Shen$^{1}$}
\affil{{\rm 1.} Department of Astronomy, University of Texas, Austin, Texas 78712, USA.}
%aindex{Hao, L.}
%aindex{Jogee, S.}
%aindex{Marinova, I.}
%aindex{Shen, J.}
\affil{{\rm 2.} Laboratoire d'Astrophysique, \'Ecole
Polytechnique F\'ed\'erale de Lausanne (EPFL), Observatoire de
Sauverny CH-1290 Versoix, Switzerland}
%aindex{Barazza, F.~D.}

\begin{abstract}
Galactic bars are the most important driver of secular evolution in
  galaxies. They can efficiently drive gas into the central kiloparsec of
  galaxies, thus feed circumnuclear starbursts, and possibly
  help to fuel AGN.  The connection between bars and AGN activities
  has been actively debated in the past two decades. Previous work
  used fairly small samples and often lacked a proper control sample. They reported conflicting results on
  the correlation between bars and AGN activity. Here we revisit the
  bar-AGN and bar-starburst connections using the analysis of bars in a
  large sample of about 2000 SDSS disk galaxies \citep*{bar_etal_08}. We find that AGN and star-forming galaxies have
  similar optical bar fractions, 47\% and 50\%, respectively. Both bar fractions are higher than that in inactive galaxies  (29\%). We discuss the implications of the study on the relationship between host galaxies and
  their central activities.
\end{abstract}

\section{Introduction}
Large-scale bars are very common in disk galaxies. 
At near infrared (NIR) wavelengths, the optical bar fraction averaged
across different Hubble types  is $\sim$~60\%  from quantitative bar identification methods, such as the ellipse fitting and structural decomposition \citep*[e.g.][]{lau_etal_04a, mar_jog_07, men_etal_07, wei_etal_09}
, and is $\sim$~72\% from
visual classification \citep{esk_etal_00}.
At optical wavelengths,  quantitative methods yield an average optical
bar fraction of  45\% to 52\%  \citep*{mar_jog_07, bar_etal_08, agu_etal_09}, while visual classification
yields  $\sim$~60\% \citep{devauc_63}. The optical bar fraction is
somewhat lower than the NIR fraction due to the obscuration of bars  by
dust lanes and star formation along the bar. 

Several studies have now moved beyond considering only the bar
fraction averaged over many Hubble types. They investigated how
the optical bar fraction varies with the Hubble types or host properties.
The optical bar fraction rises in galaxies with low Bulge-to-Disk ratio or/and 
high luminosity \citep{odewah_96, bar_etal_08, bar_etal_09, agu_etal_09, mar_etal_09}.

The non-axisymmetric stellar bar can drive gas from the outer disk 
to the central kiloparsec, where they trigger star formation. 
This is supported by several observations, which show that barred
galaxies host high gas densities and circumnuclear starbursts 
\citep*[e.g.][]{jog_etal_05} and that the central gas
concentration is larger in barred than unbarred galaxies  
\citep*[e.g.][]{sak_etal_99, she_etal_04}. Due to their efficiency in driving gas inflows in the disks, bars are strong candidates for the triggering of nuclear activities.

There is strong evidence for a connection between large-scale bars and circumnuclear starbursts. Barred galaxies show enhanced radio continuum and infrared emissions compared with unbarred ones \citep[e.g.][]{hummel_81, haw_etal_86}, and starburst galaxies tend to be more barred compared with the non-starburst galaxies \citep*[e.g.][]{arsena_89, hua_etal_96, ho_etal_97, hun_mal_99}. Bars can also set up resonances, such as the inner Lindblad resonances (ILRs), which can prevent the gas from going further in. Therefore, gas often builds up on a ring (a few hundred parsecs in radius) and the circumnuclear starbursts can occur there \citep[e.g.][]{jog_etal_05}.

The connection between bars and AGNs is less clear (see \citealt{jogee_06} for a 
review). Over the last two decades a large number of studies were carried out to identify if such a correlation exists. Most of them compared the bar fraction of the AGN sample with that of a control sample of inactive galaxies. The results are controversial. For example, studies like \citet{ho_etal_97}, \citet{hun_mal_99}, \citet{mul_reg_97}, and \citet{mar_etal_03} found no excess of bars in Seyfert galaxies, while \citet*{kna_etal_00}, \citet{hao_lai_etal_02}, and \citet*{lau_etal_04b} reported a higher fraction of bars in Seyferts. 

%The large-scale bar is only efficient in driving gas inwards to a few hundred parsec scale, so the large-scale bar may not be directly associated with the fueling of AGN. However, the correlation between the presence of large-scale bar and AGN may still exist, if the mechanisms that are responsible for the fueling of AGNs in the inner few pc scale are favored by the existence of a large-scale bar. 

Previous studies are limited in several aspects. 
Firstly, the sizes of the samples are often small, including only a few tens of AGNs and control galaxies.
In some cases, the control sample was not well matched to the active sample.
Secondly, identifications method for bars are not always consistent across
samples.
Thirdly, the spectral classifications of galaxies as AGNs or inactive galaxies are significantly inconsistent. Most studies adopted the galaxy classifications in NASA/IPAC NED, which could be done by different people using different criteria. Our study tries to overcome
some of  these limitations by systematically investigating the 
optical bar fraction of AGNs and non-AGNs in a large number of 
galaxies from the Sloan Digital Sky Survey (SDSS), using matched
active and control samples, and the same consistent quantitative
method for identifying bars across samples. 

\section{The Sample and Their Spectral Classifications}
Our sample is based on the one in \citet{bar_etal_08}. From the 3692 galaxies in the Sloan Digital Sky Survey (SDSS) with $18.5 < M_g < -22.0$ mag and redshift $0.01<z<0.03$, \citet{bar_etal_08} selected 1961 disk galaxies via their blue colors. They applied ellipse fitting to find and characterize bars in 
these disk galaxies. They exclude 169 disk galaxies 
with failed or messy ellipse fittings, or ambiguous classifications
from the final analysis. 648 galaxies were classified as  
highly-inclined galaxies ($i>60^\circ$) and disregarded as 
morphological analysis is unreliable for such systems. In the  
final sample of 1144 moderately inclined galaxies, 
553 were barred galaxies, and 591 were unbarred galaxies. 

All the 1792 disk galaxies have corresponding SDSS spectra, which are taken with a 3\arcsec aperture, with a spectral resolution of 2200 covering a wavelength range from 3700~\AA~to 9000~\AA. The 3\arcsec aperture size corresponds to 606 pc to 1.78 kpc over $0.01<z<0.03$, therefore, our spectra and the corresponding spectral classifications are done to the circumnuclear region of a typical disk galaxy. The spectral classification of these galaxies are done with various emission lines in the SDSS spectra.  We process the spectra, measure the emission lines, and classify the galaxies following \citet{hao_etal_05_agn1}. In particular, we measure the emission lines after removing the stellar absorption using a set of well-developed absorption-line templates. Galaxies with weak or no emission lines, defined specifically by having EW(H$\alpha$) $< 3$\AA,  are selected first. Such galaxies have little gas and are considered ``inactive'' in our study. The rest have strong enough emission lines indicating some activities in their nuclei. Galaxies with broad H$\alpha$ emission lines (${\rm FWHM(H\alpha)}>1200$ km/s) are automatically classified as AGNs, as broad emission lines are distinctive features of Seyfert~1 like AGNs. In our sample, there are 6 broad-line AGNs. For the remaining emission line galaxies, we classify them using the BPT diagram \citep*{bal_etal_81}.  

\begin{figure}[htb]
\centerline{ \includegraphics[angle=0,width=.6\hsize]{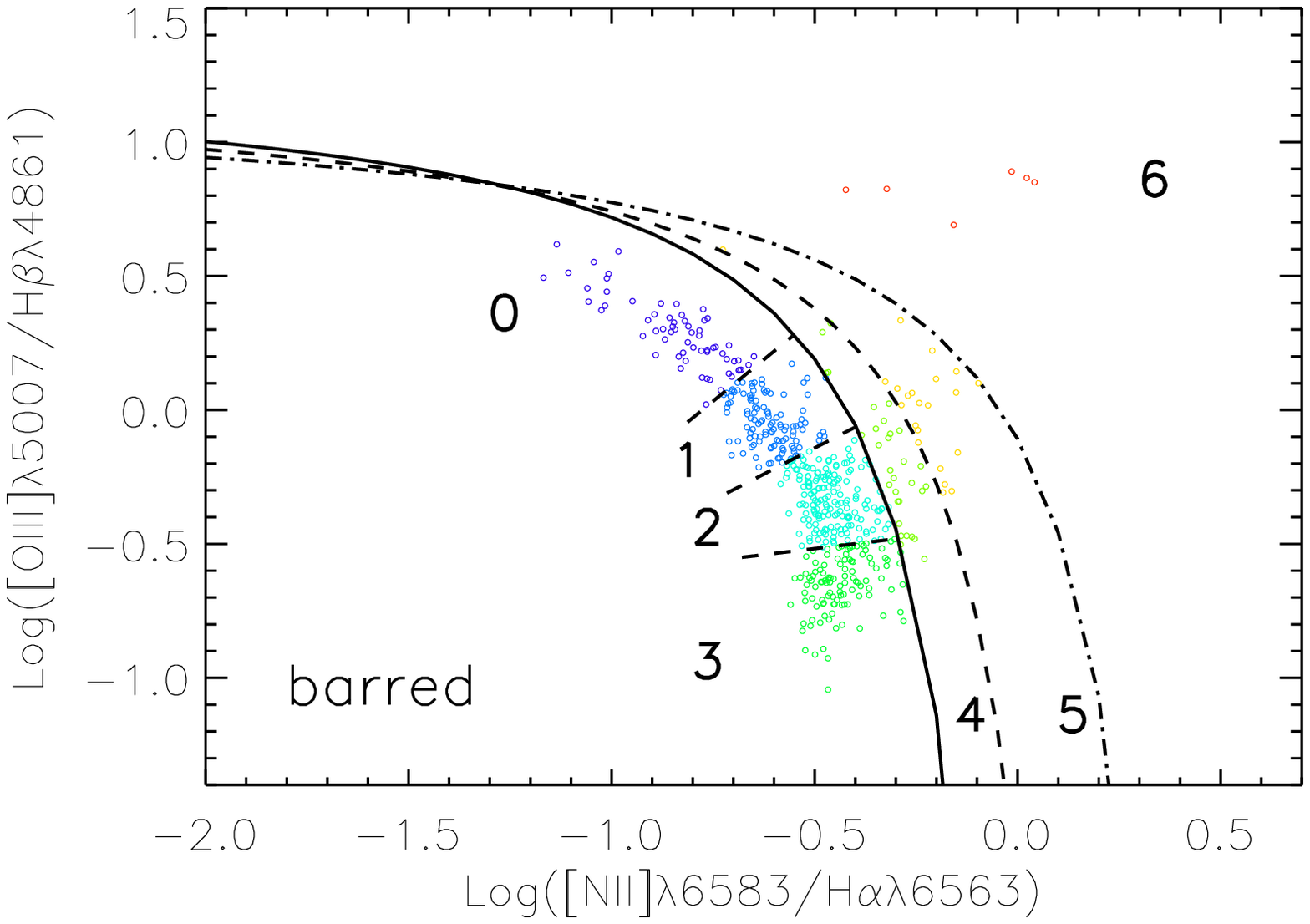}}
\centerline{ \includegraphics[angle=0,width=.6\hsize]{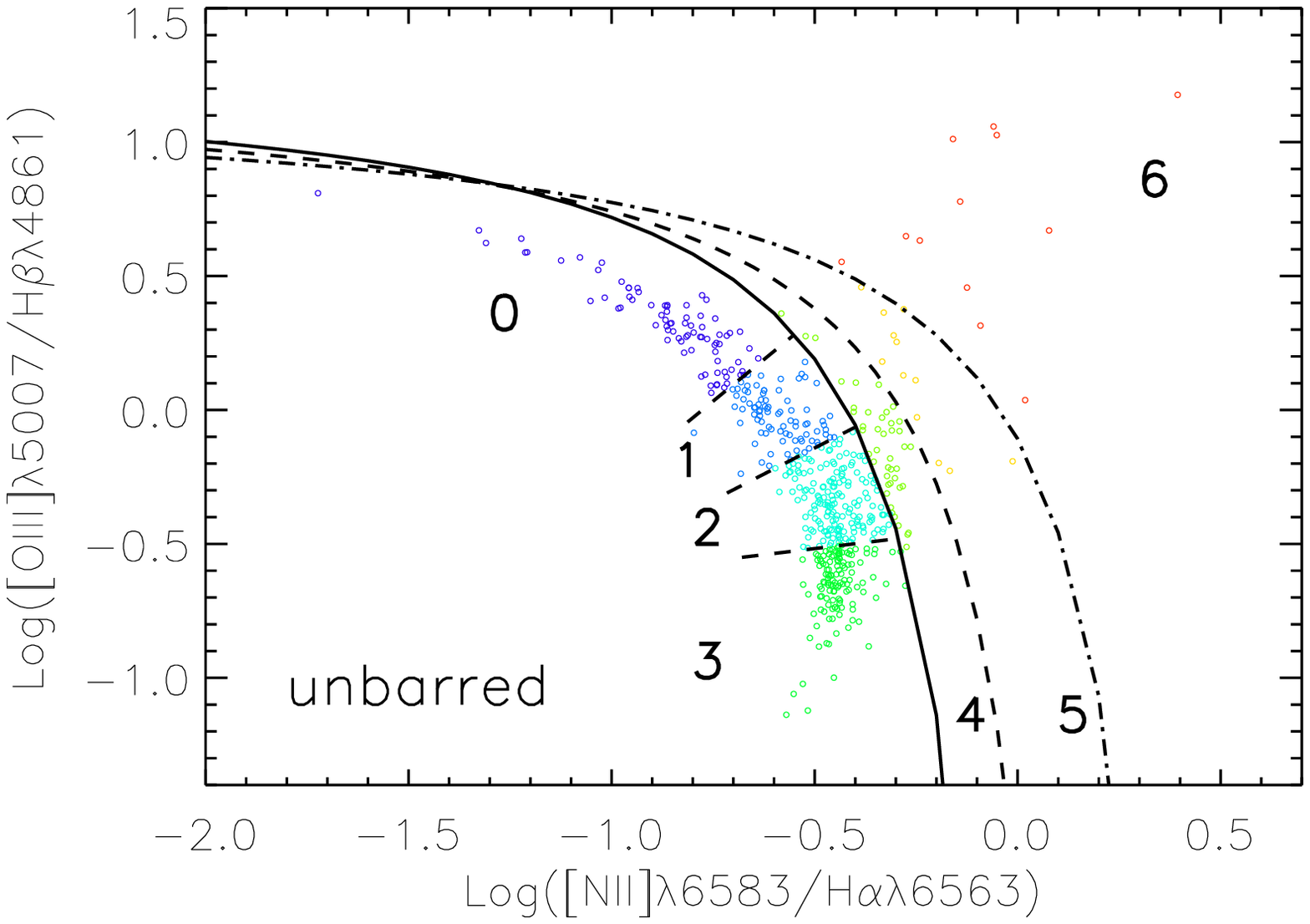}}
\caption{The BPT diagram for optically barred (upper) and optically unbarred (lower) galaxies in our sample. The solid line is taken from \citet{kau_etal_03}, and the dot-dashed line is taken from \citet{kew_etal_01}. We divide galaxies into seven spectral classes (indicated by dashed, solid, and dot-dashed lines) by their locations on the diagram, marked with the numbers.
\label{bpt}}
\end{figure}

In Figure~\ref{bpt}, we plot on the BPT diagram our sample of barred
and unbarred galaxies in the upper and lower panel respectively. In the diagram, galaxies naturally distribute
in two branches, indicating different excitation mechanisms. AGNs 
locate in the right branch, with stronger AGNs typically having higher
[NII]/H$\alpha$ and [OIII]/H$\beta$ ratios. Galaxies with pure stellar
excitation are located in the left branch, and those with lower
metallicities have higher [OIII]/H$\beta$ but lower [NII]/H$\alpha$
ratios.  The solid line, empirically defined by
\citet{kau_etal_03}, separates the two branches. These authors
classify galaxies below the line as star forming galaxies, and those
above it as AGNs. The dot-dashed line is taken from
\citet{kew_etal_01}, and demarcates the maximum position that can be
obtained by pure photo-ionization models. Galaxies located above the
line require an additional excitation mechanism, such as an AGN, or
strong shocks. The general classification scheme using the two
separation lines considers objects above Kewley's
line as AGN dominated sources, between Kewley's and Kauffmann's line
as composite AGN and starburst sources, and below the Kauffmann's line
as star forming galaxies.

Since the locations of galaxies on the BPT diagram
broadly reflect the properties of their nuclear activities, such as the
dominance of the AGN component or the metallicity of the pure
stellar-excited galaxies, we further divide galaxies into spectral
classes of 0 to 6 based on their locations on the diagram (as shown in
Figure 1). We would like to investigate whether the optical bar fraction changes
with these nuclear properties. Galaxies with spectral classes of 0 to 3 are broadly
considered as star-forming galaxies, and 4 to 6 as AGNs. We assign
inactive galaxies, which have little emission lines as spectral
class of $-1$.

We check the distributions of various host galaxy properties of galaxies with different spectra classifications, such as the redshift range, the $r$-band absolute magnitude, the stellar mass, and the sersic index. We found no significant differences of the host galaxy properties between inactive galaxies and star-forming galaxies or active galaxies. 

\section{Results}

\begin{figure}[htb]
\centerline{ \includegraphics[angle=0,width=.6\hsize]{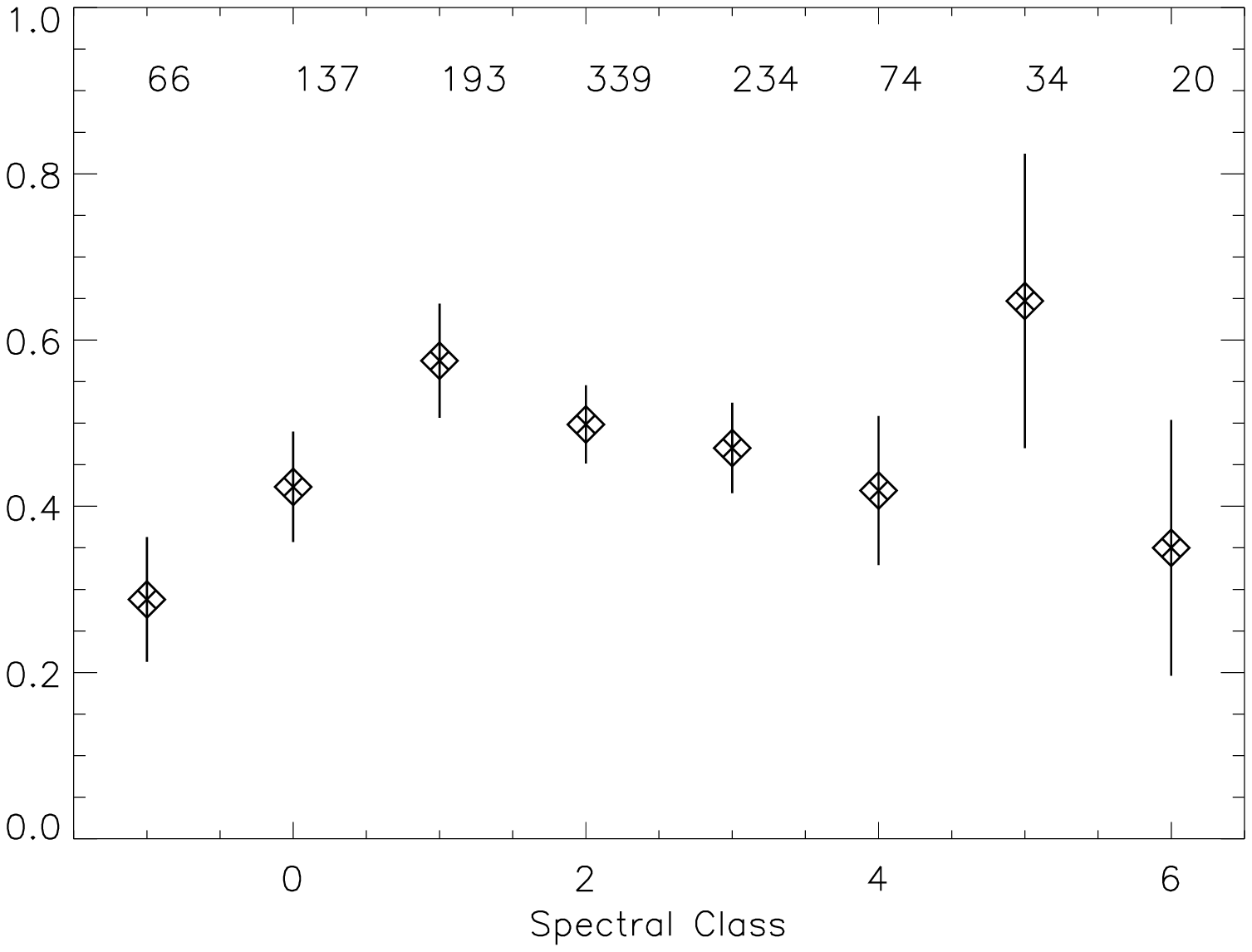}}
\centerline{ \includegraphics[angle=0,width=.6\hsize]{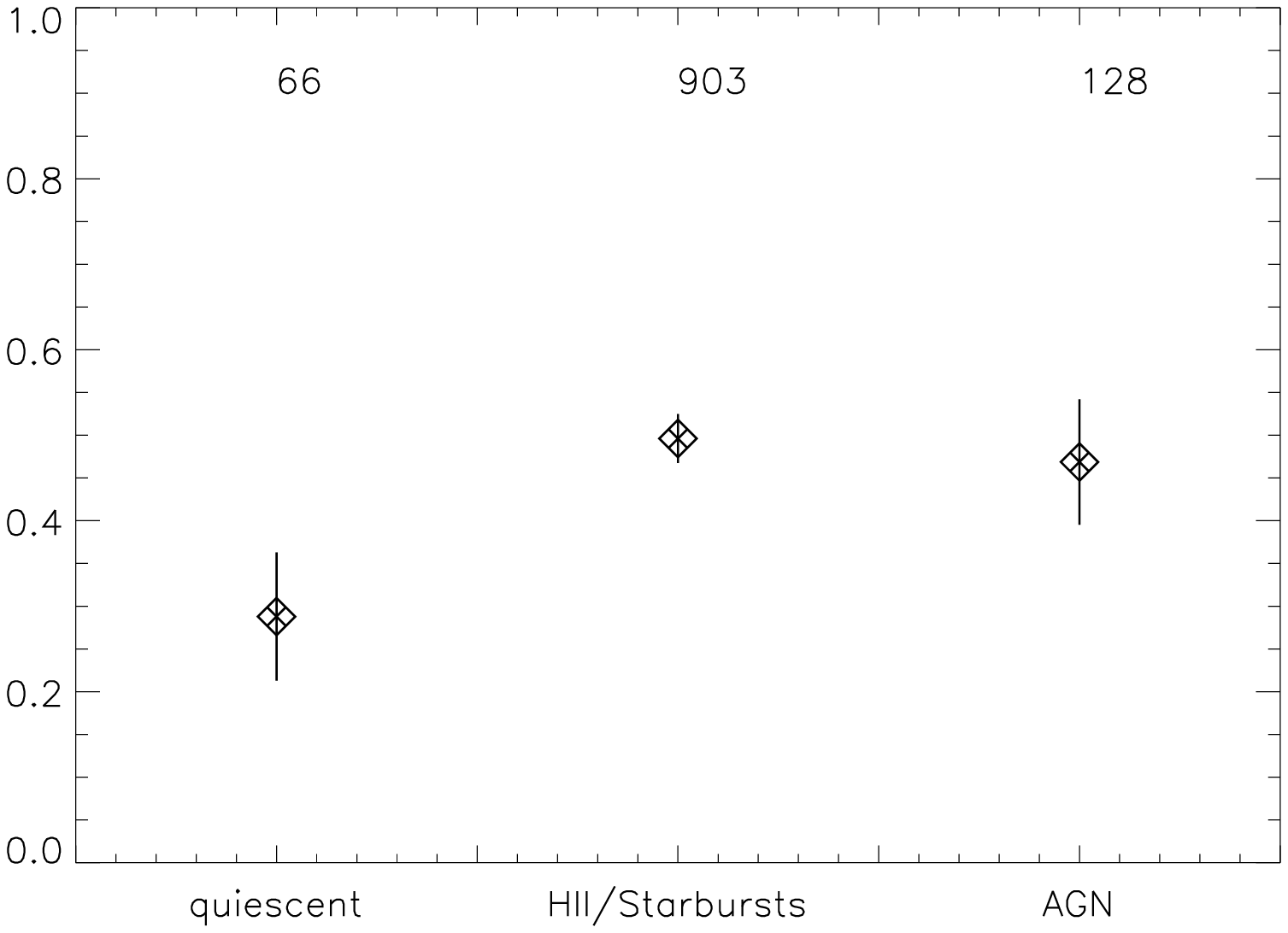}}
\caption{The optical bar fractions as a function of spectral classes defined in Figure~\ref{bpt} are shown in the upper panel. Galaxies with spectral class of $-1$ are inactive galaxies, $0$ to $3$ can be considered as star-forming galaxies, and $4$ to $6$ as AGNs. In the lower panel, we show the optical bar fraction for galaxies of the three broad classifications. The numbers at the top of each figure are the total number of moderately inclined disk galaxies in each spectral class.
\label{barf1}}
\end{figure}

In Figure~\ref{bpt}, we find no clear differences between the
distributions of barred and unbarred galaxies on the BPT
diagram. Another way to look at it is shown in the upper panel of Figure~\ref{barf1}, where we plot the optical bar fraction of galaxies with different
spectral classifications ($-1$ to $6$) defined by their locations on
the BPT diagram (see Section 3). We find that inactive galaxies have the lowest optical bar
fraction with only 29\%. Galaxies with other spectral classes have similar bar
fraction within the error bars. This is more obvious when we combine
the spectral class of 0 to 3 as starbursts and 4 to 6 as AGNs. The optical bar
fraction of inactive galaxies, starburst galaxies, and AGNs are
shown in the lower panel of Figure~\ref{barf1}. We find that the optical bar
fractions of AGNs and starburst galaxies are similar, at 50\% and 47\%
respectively, both are about $\sim~1.6$ times higher than the optical bar fraction of the
inactive galaxies. 

Our result suggests that AGNs have an excess optical bar fraction compared with the inactive galaxies, but show no excess compared with the starburst galaxies. Therefore accurate and consistent spectroscopic classification of both the AGN sample and the control sample is important in evaluating the excess of bars in AGNs. Many previous studies have overlooked this issue. Among three studies \citep{ho_etal_97, hun_mal_99, lau_etal_04b} where we can clearly decide the dominant spectral classes of the comparing sample, our result agrees with two of them. The comparing sample in \citet{ho_etal_97} is mainly composed of star-forming galaxies and they found no excess optical bar fraction in AGNs, which agrees with our result. Based on the classification in NED, \citet{lau_etal_04b} divide galaxies into Seyferts, LINERs, starbursts, and inactive galaxies. They found a similar NIR bar fraction for Seyfert galaxies, LINERS, and HII/starburst galaxies at 72\%, compared to 55\% in nonactive galaxies. The pattern also agrees with our result. The absolute values of the NIR bar fraction in \citet{lau_etal_04b} are 
higher than our optical bar fraction. This could be due to two factors. 
Firstly, the NIR bar fraction is known to be higher than the 
optical one by a factor of $\sim$ 1.3 (see $\S$ 1), due to the 
obscuration of bars by dust and star formation. 
Secondly, the number of barred galaxies could be underestimated 
%%% by about 7\% in \citet{bar_etal_08}, 
by a factor of $1.14$ in \citet{bar_etal_08}, as they regard 
galaxies with twisted position angles, but otherwise bar-like 
features to be unbarred galaxies. These galaxies could be 
weakly barred galaxies. 

Our result however, disagrees with \citet{hun_mal_99}, who 
found that the Seyfert and LINERs in the Extended 12 $\mu$m 
Galaxy Sample (E12GS) have an optical 
bar fraction of 68\% and 61\%, 
similar to the inactive galaxies in the E12GS at 69\%. 
Star-forming galaxies in their sample have a higher 
optical bar fraction (85\%) than both inactive 
galaxies and Seyferts.  Our disagreement could be due to two factors. Firstly, \citet{hun_mal_99} used RC3, which is the visual classification to identify bars. Therefore, their optical bar fractions are higher than ours where bars are identified by ellipse fitting (see $\S$ 1). Secondly, the spectral classifications of galaxies in \citet{hun_mal_99} are adopted from NED. Galaxies with composite AGN and star-forming contributions can easily be mis-identified.  

\begin{figure}[htb]
\centerline{ \includegraphics[angle=0,width=\hsize]{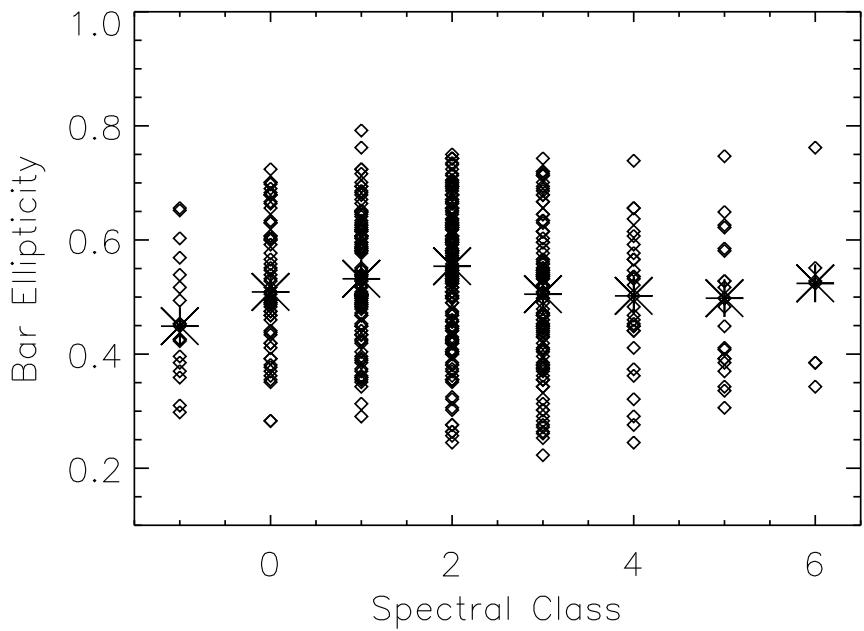}}
 \caption{The bar ellipticity of optically-visible bars as a function of the spectral classes defined in Figure~\ref{bpt}. The big asterisks are the mean ellipticity of galaxies in each spectral class.
\label{bell}}
\end{figure}

From the ellipse fitting, \citet{bar_etal_08} also obtained the ellipticity of optically visible bars. In Figure 3, we show the bar ellipticity of barred galaxies in different spectral classes. The value of the bar ellipticity varies widely for every spectral class. We plot the mean of the bar ellipticity of each spectral class with big asterisk and find it does not change with spectral classes. In particular, galaxies with stronger AGN component (from spectral 
class 4 to 6) do not show weaker bar strengths. 
Therefore, we find no indication of bar weakening by AGNs. This is 
consistent with the theoretically robustness of bars \citep*[e.g.][]{she_sel_04, ath_etal_05}.
They estimate 
that central mass concentrations in the form of super-massive black 
holes in present-day galaxies  fall well below the limit to significantly
weaken bars.

\section{Conclusions}

With the classification and structural information of $\sim 2000$ disk galaxies from the SDSS \citep{bar_etal_08}, we study the optical bar fraction of AGNs, star-forming, and inactive galaxies from the sample. We find that the optical bar fraction of the AGNs is 47\%, similar to the optical bar fraction of the star-forming galaxies (50\%). Both are higher than the optical bar fraction of the inactive galaxies (29\%). This suggests that accurate and consisitent spectral classification is important in evaluating whether there is an excess of bars in AGNs, and could be the reason for controversial results reported in previous studies on the issue. Our study has several imporvements compared to previous ones. The size of our sample is large. We have consistent SDSS spectra for all our galaxies, therefore, we can obtain accurate and consistent spectral classifications for the sample. In addition, the SDSS spectra are taken with the 3\arcsec aperture, which corresponds to 606 pc to 1.78 kpc in the redshift range of $0.01<z<0.03$. This scale matches well with the typical circumnuclear region of a galaxy, therefore, the spectra are perfect at probing the circumnuclear stubursts. 

We find an excess of the optical bar fraction of star-forming galaxies and AGNs compared to the inactive galaxies. This suggests that large-scale primary bars drive gas to inner kpc where they pile up near the ILRs, fueling  circumnuclear starbursts. The gas pile up is in someway also related with the increase of the AGN activity. But to feed the AGN directly, the
  gas in the inner kpc still has to reduce its angular momentum by several
  orders of magnitude  (see Figure 3 in \citealt{jogee_06}) and a secondary mechanism
  is then needed to drive the gas further in  (e.g., nuclear bars,
  dynamical friction). The latter may not be necessarily coupled to
  the primary bar. This agrees with our result that we do not observe an excess of bar fraction of AGNs compared to the star-forming galaxies. Furthermore, we find no evidence of bar weakening by AGNs. This agrees with previous theoretical expectations \citep{she_sel_04, ath_etal_05}.

There is one caveat of our study. Our bar identification is based on the optical data instead of NIR, therefor our bar fraction is in general lower than the typical NIR bar fraction. However, we do not expect our comparative results to change, unless the obscuration of bars by dust and star formation have preferential effects.  

%\bibliographystyle{aj.bst}
%\bibliography{baragnref.bib}

\end{document}